\documentclass[aps,twocolumn,prl,floatfix,showpacs,superscriptaddress]{revtex4-1}

\usepackage{amsmath}
\usepackage{amssymb}
\usepackage{amsfonts}
\usepackage[pdftex]{graphicx}
\usepackage{units}
\usepackage[usenames]{color}
\usepackage{dcolumn}
\usepackage{bm}
\usepackage{float}

\renewcommand{\vec}[1]{\ensuremath{\boldsymbol{#1}}}

\begin{document}
\title{Topological Crystalline Transition Metals: Strained W, Ta, Mo, and Nb}

\author{Danny Thonig}
\affiliation{Department of Physics and Astronomy, Material Theory, University Uppsala, Box 516, 75120 Uppsala, Sweden}

\author{Tom\'{a}\v{s} Rauch}
\affiliation{Institute of Physics, Martin Luther University Halle-Wittenberg, Von-Seckendorff-Platz 1, 06120 Halle (Saale), Germany}

\author{Hossein Mirhosseini}
\altaffiliation[Present address: ]
                     {Max Planck Institute for Chemical Physics of Solids, N\"othnitzer Str. 40, 01187 Dresden , Germany}
\affiliation{Max Planck Institute of Microstructure Physics, Weinberg 2, 06120 Halle (Saale), Germany}

\author{J\"urgen Henk}
\affiliation{Institute of Physics, Martin Luther University Halle-Wittenberg, Von-Seckendorff-Platz 1, 06120 Halle (Saale), Germany}
\email[Corresponding author.]{juergen.henk@physik.uni-halle.de}

\author{Ingrid Mertig}
\affiliation{Institute of Physics, Martin Luther University Halle-Wittenberg, Von-Seckendorff-Platz 1, 06120 Halle (Saale), Germany}
\affiliation{Max Planck Institute of Microstructure Physics, Weinberg 2, 06120 Halle (Saale), Germany}

\author{Henry Wortelen}
\affiliation{Physikalisches Institut, Westf\"alische Wilhelms-Universit\"at, Wilhelm-Klemm-Str. 10, 48149 M\"unster, Germany}

\author{Bernd Engelkamp}
\affiliation{Physikalisches Institut, Westf\"alische Wilhelms-Universit\"at, Wilhelm-Klemm-Str. 10, 48149 M\"unster, Germany}

\author{Anke B. Schmidt}
\affiliation{Physikalisches Institut, Westf\"alische Wilhelms-Universit\"at, Wilhelm-Klemm-Str. 10, 48149 M\"unster, Germany}

\author{Markus Donath}
\affiliation{Physikalisches Institut, Westf\"alische Wilhelms-Universit\"at, Wilhelm-Klemm-Str. 10, 48149 M\"unster, Germany}

\date{\today}

\begin{abstract}
In a joint theoretical and experimental investigation we show that a series of transition metals with strained body-centered cubic lattice---W, Ta, Nb, and Mo---host surface states that are topologically protected by mirror symmetry. Our finding extends the class of topologically nontrivial systems by topological crystalline transition metals. The investigation is based on independent calculations of the electronic structures and of topological invariants, the results of which agree with established properties of the Dirac-type surface state in W(110). To further support our prediction, we investigate both experimentally by spin-resolved inverse photoemission and theoretically an unoccupied topologically nontrivial surface state in Ta(110).
\end{abstract}


\maketitle

\paragraph{Introduction.}
Topological insulators have become an exciting topic in condensed matter physics \cite{Hasan10}. Insulating in the bulk, these systems host topologically protected surface states that cross the global band gap and exhibit spin-momentum locking. These features render them very valuable for fundamental research and promising for spin-electronic applications. Up to now, most investigations have addressed strong topological insulators (TIs; in particular strained HgTe \cite{Koenig07} as well as Bi$_{2}$Te$_{3}$ and similar compounds \cite{Hasan11}) and topological crystalline insulators (TCIs; e.\,g.\ SnTe \cite{Fu11,Xu12}).

\begin{figure*}
\centering
 \includegraphics[width = 0.9\textwidth]{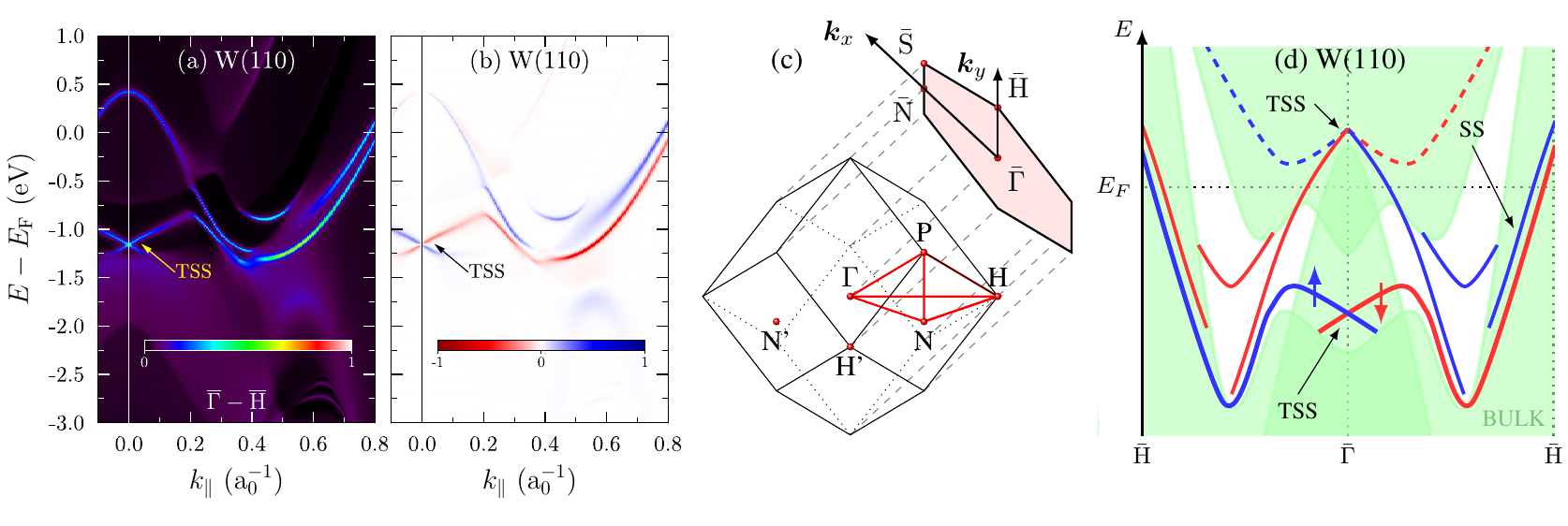}
\caption{(Color online) Surface electronic structure of W(110). (a) Topological surface state (marked TSS). The spectral density of the topmost layers is shown as normalized color scale for the $\overline{\Gamma}$--$\overline{\mathrm{H}}$ line of the surface Brillouin zone. (b) As (a), but spin-resolved with respect to the Rashba component of the spin polarization ($\pm1$ is fully spin polarized: $-1$ spin-down, $+1$ spin-up).  (c) Brillouin zone of a bcc lattice and its projection onto the (110) surface. Red lines indicate an irreducible part. $\mathrm{H}$ and $\mathrm{H}'$ become inequivalent under strain. (d) Schematic illustration of the surface-state dispersion from (a), with spin polarization indicated by colors.}
\label{fig:Schematic}
\end{figure*}

Recent experimental investigations of the surface electronic structure of W(110) have brought a remarkable surface state to attention \cite{Miyamoto12,Miyamoto12c,Hochstrasser02}, followed up by theoretical calculations \cite{Rybkin12,Giebels13,Mirhosseini13,Mirhosseini13b,Braun14}. This surface state is strongly spin-polarized due to Rashba spin-orbit coupling \cite{Bychkov84b,Bychkov84a,Winkler03}. But more strikingly, it shows linear and strong dispersion along the $\overline{\Gamma}$--$\overline{\mathrm{H}}$ high-symmetry line of the surface Brillouin zone (Fig.~\ref{fig:Schematic}a). It is therefore reminiscent of a TI's surface state \cite{Hasan10,Hasan11}. While these Dirac states possess a cone-like dispersion, their counterpart at W(110) becomes flattened along $\overline{\Gamma}$--$\overline{\mathrm{N}}$ due to the two-fold rotational symmetry of the surface.

The salient properties of this surface state immediately raise the question whether it is a `true' topologically nontrivial surface state (TSS) or it is `only' reminiscent of a TSS\@. In this Letter we prove that the surface state of slightly strained (compressed by $4\%$) W(110) is indeed a TSS by calculating the respective topological invariants. On top of this, we show that it has counterparts in other transition metals with body-centered-cubic (bcc) lattice: Ta, Nb, and Mo. We provide both experimental and theoretical evidence for the exemplary material Ta. Having identified a number of topological crystalline transition metals, our findings call for investigating other material classes---besides insulators and semimetals \cite{Eremeev12,Bianchi10,Yang15,Wang15,Sun15,Yan13,Li15}---that may host topologically protected edge states. Recently, Au(111) with increased strength of the spin-orbit coupling has been identified topologically nontrivial \cite{Yan15}.

\paragraph{Theoretical aspects.} 
Electronic-structure calculations for both bulk and (110) surfaces of W, Ta, Nb, and Mo have been performed within the local density approximation to density-functional theory (DFT), using Perdew-Burke-Ernzerhof generalized gradient exchange-correlation functionals \cite{Perdew96,Perdew97}\footnote{For details, see the online Supplementary Material.}. We have applied relativistic multiple-scattering theory as formulated in the Korringa-Kohn-Rostoker (KKR) approach \cite{Henk02c,Zabloudil05}. Spin-orbit coupling is accounted for in a non-perturbative manner by solving the Dirac equation. Finite-size effects are avoided by modeling the surfaces in a semi-infinite geometry.

The KKR calculations are complemented by analogous computations with the \textsc{vasp} program package \cite{Kresse96b,Kresse96a}, using a slab geometry. The electronic structures obtained by these independent methods agree very well, putting our findings on firm ground.

Surface relaxations have been determined by \textsc{vasp} calculations (Table~\ref{tab:geometry}). They are important ingredients in our reasoning. Experimental data for W(110) agree with their theoretical counterparts: Ref.~\onlinecite{Venus00} gives $d_{12} = \unit[-2.2\pm 1.0]{\%}$, whereas Ref.~\onlinecite{Meyerheim01a} states $d_{12} = \unit[-2.7(5)]{\%}$ and $d_{23} < \unit[0.3]{\%}$.

\begin{table}
\caption{Geometry of bcc(110) surfaces obtained from \textsc{vasp} calculations. The relative changes of the distances $d_{ij}$ between layer $i$ and $j$ is given with respect to the bulk interlayer distance; $i, j = 1, 2, 3, \ldots$ indicate the topmost, second, third layer etc., $a$ is the lattice constant. Data for Ta reproduced from Refs.~\onlinecite{Engelkamp15,Wortelen15b}.}
\begin{tabular}{c|rrrr}
 \hline\hline
            & \multicolumn{1}{c}{W} & \multicolumn{1}{c}{Ta} & \multicolumn{1}{c}{Mo} & \multicolumn{1}{c}{Nb} \\
   \hline 
   $d_{12}$   & $\unit[-3.67]{\%}$    & $\unit[-4.81]{\%}$ & $\unit[-4.96]{\%}$ & $\unit[-3.77]{\%}$ \\
   $d_{23}$   & $\unit[+0.92]{\%}$    & $\unit[+0.57]{\%}$ & $\unit[+1.32]{\%}$ & $\unit[+1.19]{\%}$ \\
   $d_{34}$   & $\unit[+0.20]{\%}$    & $\unit[+0.29]{\%}$ & $\unit[+0.41]{\%}$ & $\unit[+0.11]{\%}$ \\
  \hline
   $a$        & $\unit[3.172]{\AA}$   & $\unit[3.308]{\AA}$ & $\unit[3.151]{\AA}$ & $\unit[3.323]{\AA}$ \\
 \hline\hline
\end{tabular}
 \label{tab:geometry}
\end{table}

The DFT results serve as input for tight-binding (TB) parameterizations for W, Ta, Nb, and Mo, from which we calculate Berry curvatures and mirror Chern numbers $n_{\mathrm{m}}$. A mirror Chern number classifies systems in which the topological protection is brought about by mirror symmetry (topological crystalline insulators, TCIs). We consider the mirror planes associated with the $\overline{\Gamma}$--$\overline{\mathrm{H}}$ and $\overline{\Gamma}$--$\overline{\mathrm{N}}$ lines of the surface Brillouin zone (Fig.~\ref{fig:Schematic}c). The modulus of $n_{\mathrm{m}}$ equals the number of TSSs for the respective mirror plane, its sign determines the spin chirality of these TSSs\@. The tight-binding approach proved successful for, e.\,g., Bi$_{2}$Te$_{3}$, Bi$_{2}$Se$_{3}$, SnTe, and HgTe$_x$S$_{1-x}$ \cite{Rauch14,Barone13,Rauch15}.

\paragraph{Experimental aspects.} 
Because the Dirac-type surface state of W(110) has been investigated in very detail \cite{Miyamoto12,Miyamoto12c,Miyamoto15,Mirhosseini13,Braun14,Miyamoto16}, we provide experimental evidence for TSSs in Ta(110). For this purpose we used spin- and angle-resolved inverse photoemission (IPE). The spin-dependent unoccupied electronic structure of Ta(110) was investigated by utilizing a spin-polarized electron beam \cite{Stolwijk14} and measuring the Rashba component of the spin polarization. The emitted photons with an energy of $\unit[9.9]{eV}$ are detected by Geiger-M\"{u}ller counters. The total energy resolution of the IPE experiment is about $\unit[350]{meV}$. A detailed description of the IPE experiment and in particular for the experiment on Ta(110) is given in Refs.~\onlinecite{Budke07,Wortelen15b}, respectively.

\paragraph{Results and discussion.}
Topologically nontrivial systems are characterized by band inversions which give rise to nonzero topological invariants. If the system exhibits either a global or a semimetal gap~\footnote{A \emph{global} gap separates two bands for all wave vectors $\vec{k}$; there is a $\vec{k}$-independent energy range which separates the upper from the lower band. In a \emph{semimetal} gap there is a $\vec{k}$-dependent energy range separating the two bands.} the invariant is integer, which is obviously the case for insulators. Focusing first on W(110), we are facing two problems: semimetal band gap and band inversion.

\textit{--- Semimetal band gap.} W does not have a band gap in the energy region in which the relevant surface state shows up (Fig.~\ref{fig:Schematic}a). However, both compressive or tensile strain in $[110]$ direction open up the desired semimetal gap in the $\overline{\Gamma}$--$\overline{\mathrm{H}}$ and $\overline{\Gamma}$--$\overline{\mathrm{N}}$ mirror planes, thus, allowing to compute the mirror Chern numbers $n_{\mathrm{m}}$. We have applied strain up to $\pm\unit[4]{\%}$ which is in the range of the surface relaxation (Table~\ref{tab:geometry}).

\begin{figure*}
\centering
\includegraphics[width = 0.8\textwidth]{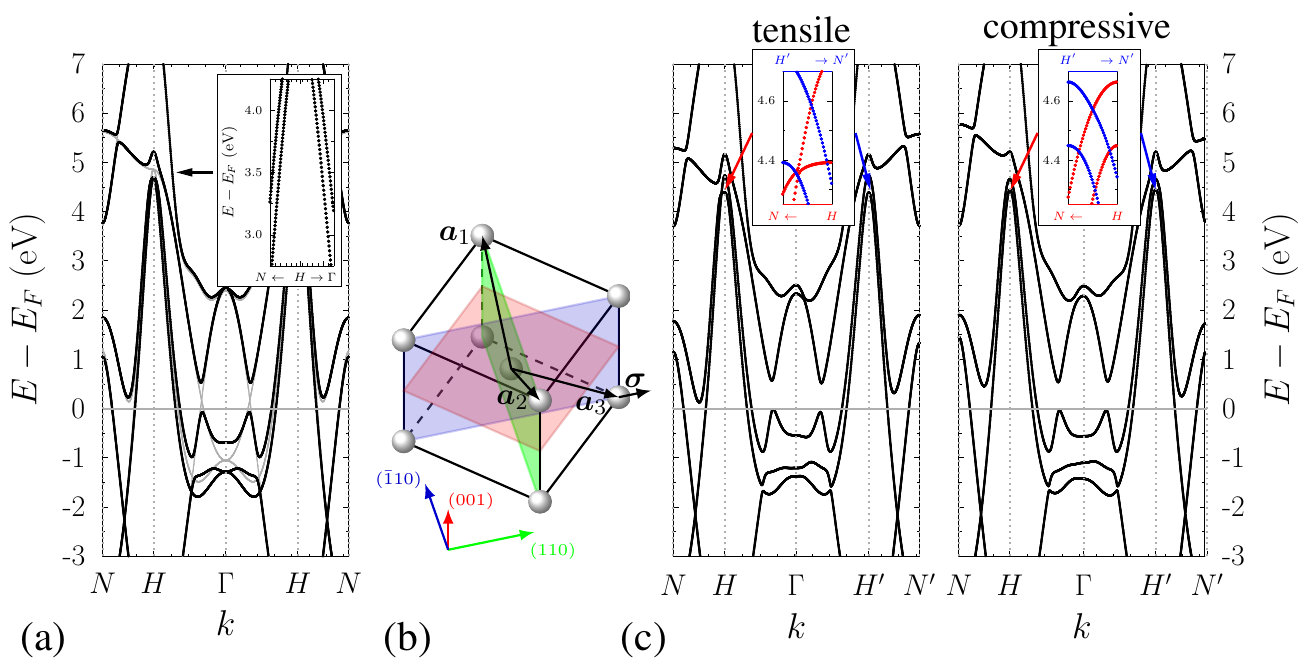}
\caption{(Color online) Band inversion and band gap opening in W\@. Tight-binding band structures are shown along those lines in the bulk Brillouin zone that are relevant for the topological phase transition. (a) Cubic W without (gray) and with (black) spin-orbit coupling. The inset shows a zoom to the topological `hotspot' of the dispersion (b) Body-centered cubic crystal, with lattice sites represented by spheres; the green area visualizes the (110) unit cell spanned by the lattice vectors $\vec{a}_1$ and $\vec{a}_2$. The lattice vector between the $(110)$ planes is $\vec{a}_3$ which is altered upon strain $\vec{\sigma}$. The two mirror planes --- with surface normal in $(001)$ and $(\bar{1}10)$ --- for which the Chern numbers have been computed are displayed in blue and red. (c) Tensile and compressively strained W with spin-orbit coupling. Insets show band dispersions that are relevant for the topology of W\@ under strain at $\mathrm{H}$ (red) and $\mathrm{H}'$ (blue). $\mathrm{H}'$ and  $\mathrm{N}'$ are defined in Fig.~\ref{fig:Schematic}c.}
\label{fig:Band-inversion}
\end{figure*}

\textit{--- Band inversion.} Although the Dirac-type surface state is observed at the $\overline{\Gamma}$ point of the surface Brillouin zone, the relevant band inversion takes place at the $\mathrm{H}$ points of the bulk Brillouin zone (Fig.~\ref{fig:Band-inversion}), i.\,e.\ at energies larger than the Fermi level $E_{\mathrm{F}}$. The `small group' of $\mathrm{H}$ is $\mathrm{O}_{\mathrm{h}}$ and the six $t_{2\mathrm{g}}$ bulk bands are  split into one twofold degenerate $E_{5/2\mathrm{g}}$ and one fourfold degenerate $G_{3/2\mathrm{g}}$ level if spin-orbit coupling is considered. Under strain along $[110]$, the `small group' of $\mathrm{H}$ is reduced to $\mathrm{D}_{2\mathrm{h}}$. The $E_{5/2\mathrm{g}}$ level stays twofold degenerate, the $G_{3/2\mathrm{g}}$ level is further split into two twofold degenerate levels. All these levels belong to the representation $E_{1/2\mathrm{g}}$. This splitting shows up for both tensile and compressive strain.

For the strained systems with semimetal band gaps we compute the mirror Chern numbers $n_{\mathrm{m}}$ for both the $\overline{\Gamma}$--$\overline{\mathrm{H}}$ and $\overline{\Gamma}$--$\overline{\mathrm{N}}$ mirror planes. For tensile strain we find $n_{\mathrm{m}} = 0$, indicating a topological trivial system. For compressive strain, $n_{\mathrm{m}}$ for the $\overline{\Gamma}$--$\overline{\mathrm{N}}$ mirror plane vanishes as well; we recall that the surface state is weakly dispersive along this line \cite{Miyamoto12c}. However, for the $\overline{\Gamma}$--$\overline{\mathrm{H}}$ line---for which the surface state shows linear dispersion---we compute $n_{\mathrm{m}} = -2$. This finding indicates that compressively strained W is a topological crystalline metal. It further tells that the semimetal band gap has to be bridged by two TSSs with identical spin chirality. Analogous calculations for Ta, Mo, and Nb give identical results concerning the topological properties.

\begin{figure}
\centering
\includegraphics[width = 0.95\columnwidth]{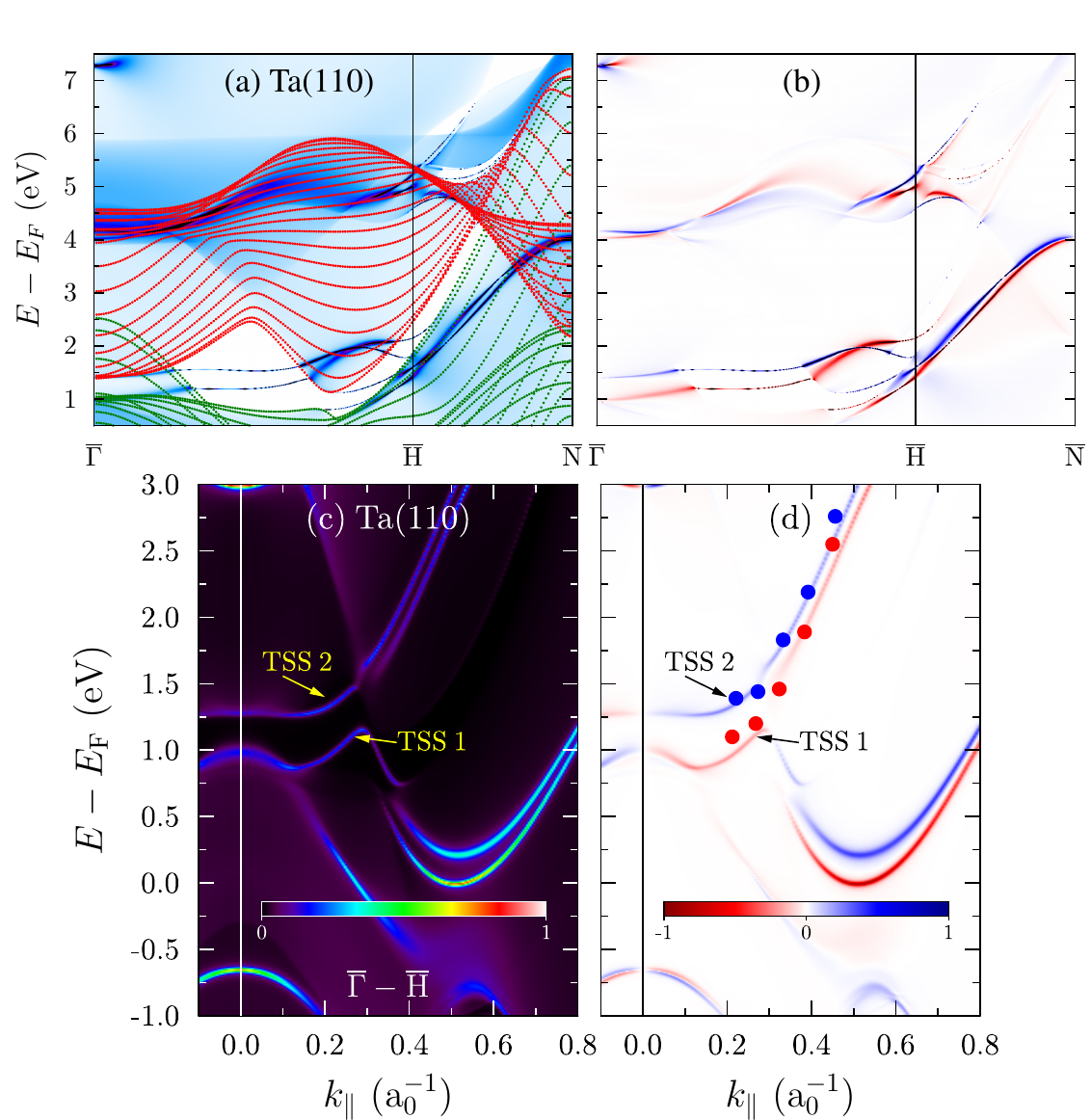}
\caption{(Color online) Unoccupied electronic structure of Ta(110) for the $\overline{\Gamma}$--$\overline{\mathrm{H}}$ line of the surface Brillouin zone. (a) Spectral density of the topmost surface layer calculated with the tight-binding method. The bulk bands that form the boundary of the $(E, \vec{k}$)-dependent band gap are shown in green and red. (b) As (a) but resolved with respect to the Rashba spin component. (c), (d) As (a) and (b) but calculated by the \textit{ab initio} KKR method. The symbols result from spin-resolved inverse-photoemission experiments and indicate peak positions derived from
spectra shown in the Supplementary Material. The topological surface states are marked TSS1 and TSS2.}
\label{fig:Ta}
\end{figure}

To provide qualitative insight into the complicated electronic structure we turn to Ta(110) (Fig.~\ref{fig:Ta}a). In the TB calculations for the surface we assume homogeneously strained samples (i.\,e., without detailed surface relaxation) and bulk TB parameters in the topmost layers. By calculating the bulk band structure along $\overline{\Gamma}$--$\overline{\mathrm{H}}$--$\overline{\mathrm{N}}$ for a set of equidistant $k_{\perp}$, we achieve a representation of the $(E, \vec{k})$-dependent band gap: the band that forms its lower (upper) boundary is colored green (red). This band structure is superimposed onto the surface spectral density which shows two surfaces states. The surface state TSS1 starts at $\unit[1.0]{eV}$ at $\overline{\Gamma}$ off a green bulk band and can be traced to $\overline{\mathrm{H}}$ where it snuggles up to bulk band edge; then it disperses to higher energies at $\overline{\mathrm{N}}$ where it connects to a red bulk band. TSS2 starts at $\unit[1.45]{eV}$ at $\overline{\Gamma}$ off a red band and reaches a green bulk band close to $\overline{\mathrm{H}}$. The spin-resolved spectral density tells that TSS1 and TSS2 have opposite spin polarization (Fig.~\ref{fig:Ta}b). Note that TSS1 (TSS2) exhibits its part above (below) its Dirac point at $\overline{\Gamma}$ which has opposite spin polarization as compared to its lower (upper) part (cf.\ Fig.~\ref{fig:Schematic}b and d for W). The spin chirality of TSS1 and TSS2 is therefore identical, which is in line with the mirror Chern number.

We now show by \textit{ab initio} calculations and IPE experiments that Ta(110) also hosts TSSs. Ta lends itself for an investigation because its strong spin-orbit coupling produces a sizable spin-orbit band gap. The semimetal gap is between the dispersive bands that become inverted by compressive strain [cf.\ Fig.~2 of the Supplementary Material]. Near $\overline{\Gamma}$ it shows up at $E_{\mathrm{F}} + \unit[1.1]{eV}$ (Fig.~\ref{fig:Ta}c). Two surface bands with opposite spin polarization are split off the bulk band edges at $\overline{\Gamma}$, one from the lower, the other from the upper band edge, in accordance with $n_{\mathrm{m}} = -2$ (\ Figs.~\ref{fig:Ta}c and~d as well as Fig.~\ref{fig:Schematic}d). The two TSSs do not exhibit the typical Rashba-type dispersion, as observed in Au(111) and Bi/Ag(111) \cite{LaShell96,Ast07a,Requist15}. Along the $\overline{\Gamma}$--$\overline{\mathrm{H}}$ line they disperse in `unison' with nonlinear dispersion~\footnote{The linear dispersion in W(110) is explained by charge transfer \cite{Braun14}.}. Surface bands appear at lower energies relative to the bulk band edges than those of W, which is attributed to the larger lattice constant $a$ and the stronger surface relaxation of Ta compared to W (Table~\ref{tab:geometry}). An increased lattice constant `flattens' the bulk bands, resulting in down-shifted surface bands and stronger hybridizations with the bulk states. Since the TSS in W(110) is located at the lower boundary of the band gap at $\overline{\Gamma}$ (about $\unit[-1.25]{eV}$ to $\unit[-0.75]{eV}$ in Fig.~\ref{fig:Schematic}a; Ref.~\onlinecite{Mirhosseini13}), the respective Ta surface state and its Dirac point are `hidden' in the bulk bands at $\overline{\Gamma}$ \cite{Wang11,Franz13}.

\begin{figure}
\centering
\includegraphics[width = 0.95\columnwidth]{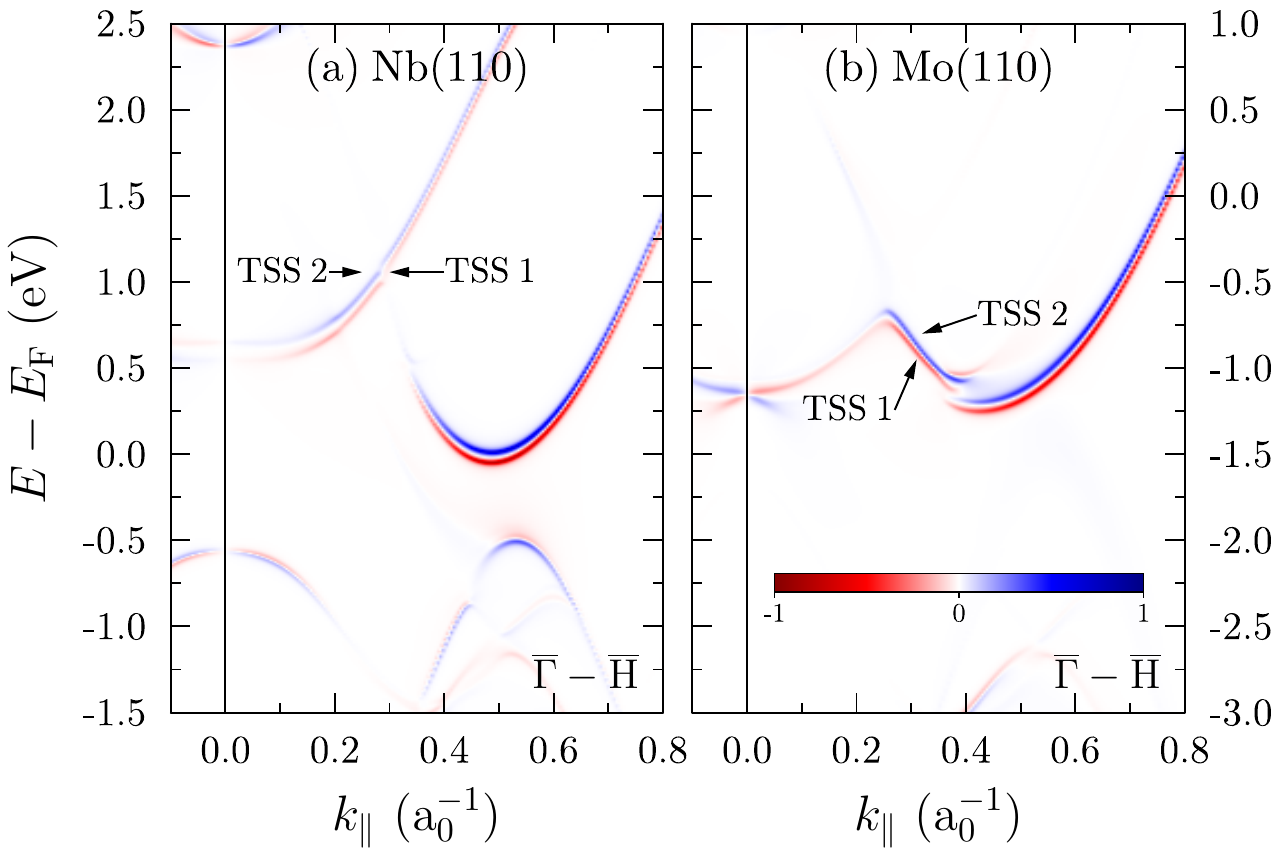}
\caption{(Color online) Spin-resolved surface electronic structure of Nb(110) (a) and Mo(110) (b), analogous to Fig.~\ref{fig:Ta}d.}
\label{fig:Nb-and-Mo}
\end{figure}

Strained Ta, Nb, and Mo possess the same topological invariants as W\@. Both Nb and Mo host TSSs as well (Fig.~\ref{fig:Nb-and-Mo}; cf.\ Ref.~\onlinecite{Rotenberg99} for Mo). Because both Nb ($Z = 41$, $4d^{4}5s^{1}$) and Mo ($Z = 42$, $4d^{5}5s^{1}$) are lighter than Ta ($Z = 73$, $5d^{3}6s^{2}$) and W ($Z = 74$, $5d^{4}6s^{2}$) the band gaps that are induced by the spin-orbit interaction are significantly smaller. The surface state in Mo resembles a Dirac-like state at $\overline{\Gamma}$ and $E-E_F = \unit[-1.2]{eV}$; the dispersion is not linear, in agreement with experiment \cite{Shikin08,Chernov15}. The orbital composition of the surface states is similar to those in Ta(110) and W(110). 

\paragraph{Concluding remarks.}
Our findings for the transition metals W, Ta, Nb, and Mo with strained bcc structure extend the class of topologically nontrivial systems by topological crystalline metals. Their (110) surfaces are effectively compressed (Table~\ref{tab:geometry}); hence, we propose that the four considered bcc metals appear topologically nontrivial at their (110) surfaces although their bulk lattice is cubic.

To confirm the predictions, we encourage investigations of Nb and Mo, preferably by spin-resolved conventional or inverse photoemission. The topological phase transition upon strain could be studied by growing metal films on either different substrates or on a piezo crystal \cite{Thiele07,Boettcher13}.

\acknowledgments
We acknowledge fruitful discussions with Albert Fert. This work is supported by the Priority Program 1666 of DFG\@.

\bibliographystyle{apsrev4-1}
\bibliography{Dirac.bcc110.bbl}

\begin{thebibliography}{55}%
\makeatletter
\providecommand \@ifxundefined [1]{%
 \@ifx{#1\undefined}
}%
\providecommand \@ifnum [1]{%
 \ifnum #1\expandafter \@firstoftwo
 \else \expandafter \@secondoftwo
 \fi
}%
\providecommand \@ifx [1]{%
 \ifx #1\expandafter \@firstoftwo
 \else \expandafter \@secondoftwo
 \fi
}%
\providecommand \natexlab [1]{#1}%
\providecommand \enquote  [1]{``#1''}%
\providecommand \bibnamefont  [1]{#1}%
\providecommand \bibfnamefont [1]{#1}%
\providecommand \citenamefont [1]{#1}%
\providecommand \href@noop [0]{\@secondoftwo}%
\providecommand \href [0]{\begingroup \@sanitize@url \@href}%
\providecommand \@href[1]{\@@startlink{#1}\@@href}%
\providecommand \@@href[1]{\endgroup#1\@@endlink}%
\providecommand \@sanitize@url [0]{\catcode `\\12\catcode `\$12\catcode
  `\&12\catcode `\#12\catcode `\^12\catcode `\_12\catcode `\%12\relax}%
\providecommand \@@startlink[1]{}%
\providecommand \@@endlink[0]{}%
\providecommand \url  [0]{\begingroup\@sanitize@url \@url }%
\providecommand \@url [1]{\endgroup\@href {#1}{\urlprefix }}%
\providecommand \urlprefix  [0]{URL }%
\providecommand \Eprint [0]{\href }%
\providecommand \doibase [0]{http://dx.doi.org/}%
\providecommand \selectlanguage [0]{\@gobble}%
\providecommand \bibinfo  [0]{\@secondoftwo}%
\providecommand \bibfield  [0]{\@secondoftwo}%
\providecommand \translation [1]{[#1]}%
\providecommand \BibitemOpen [0]{}%
\providecommand \bibitemStop [0]{}%
\providecommand \bibitemNoStop [0]{.\EOS\space}%
\providecommand \EOS [0]{\spacefactor3000\relax}%
\providecommand \BibitemShut  [1]{\csname bibitem#1\endcsname}%
\let\auto@bib@innerbib\@empty
\bibitem [{\citenamefont {Hasan}\ and\ \citenamefont {Kane}(2010)}]{Hasan10}%
  \BibitemOpen
  \bibfield  {author} {\bibinfo {author} {\bibfnamefont {M.~Z.}\ \bibnamefont
  {Hasan}}\ and\ \bibinfo {author} {\bibfnamefont {C.~L.}\ \bibnamefont
  {Kane}},\ }\href@noop {} {\bibfield  {journal} {\bibinfo  {journal} {Rev.\
  Mod.\ Phys.}\ }\textbf {\bibinfo {volume} {82}},\ \bibinfo {pages} {3045}
  (\bibinfo {year} {2010})}\BibitemShut {NoStop}%
\bibitem [{\citenamefont {K\"onig}\ \emph {et~al.}(2007)\citenamefont
  {K\"onig}, \citenamefont {Wiedmann}, \citenamefont {Br\"une}, \citenamefont
  {Roth}, \citenamefont {Buhmann}, \citenamefont {Molenkamp}, \citenamefont
  {Qi},\ and\ \citenamefont {Zhang}}]{Koenig07}%
  \BibitemOpen
  \bibfield  {author} {\bibinfo {author} {\bibfnamefont {M.}~\bibnamefont
  {K\"onig}}, \bibinfo {author} {\bibfnamefont {S.}~\bibnamefont {Wiedmann}},
  \bibinfo {author} {\bibfnamefont {C.}~\bibnamefont {Br\"une}}, \bibinfo
  {author} {\bibfnamefont {A.}~\bibnamefont {Roth}}, \bibinfo {author}
  {\bibfnamefont {H.}~\bibnamefont {Buhmann}}, \bibinfo {author} {\bibfnamefont
  {L.~W.}\ \bibnamefont {Molenkamp}}, \bibinfo {author} {\bibfnamefont {X.-L.}\
  \bibnamefont {Qi}}, \ and\ \bibinfo {author} {\bibfnamefont {S.-C.}\
  \bibnamefont {Zhang}},\ }\href@noop {} {\bibfield  {journal} {\bibinfo
  {journal} {Science}\ }\textbf {\bibinfo {volume} {318}},\ \bibinfo {pages}
  {766} (\bibinfo {year} {2007})}\BibitemShut {NoStop}%
\bibitem [{\citenamefont {Hasan}\ and\ \citenamefont {Moore}(2011)}]{Hasan11}%
  \BibitemOpen
  \bibfield  {author} {\bibinfo {author} {\bibfnamefont {M.~Z.}\ \bibnamefont
  {Hasan}}\ and\ \bibinfo {author} {\bibfnamefont {J.~E.}\ \bibnamefont
  {Moore}},\ }\href@noop {} {\bibfield  {journal} {\bibinfo  {journal} {Ann.\
  Rev.\ Cond.\ Mat.\ Phys.}\ }\textbf {\bibinfo {volume} {2}},\ \bibinfo
  {pages} {55} (\bibinfo {year} {2011})}\BibitemShut {NoStop}%
\bibitem [{\citenamefont {Fu}(2011)}]{Fu11}%
  \BibitemOpen
  \bibfield  {author} {\bibinfo {author} {\bibfnamefont {L.}~\bibnamefont
  {Fu}},\ }\href@noop {} {\bibfield  {journal} {\bibinfo  {journal} {Phys.\
  Rev.\ Lett.}\ }\textbf {\bibinfo {volume} {106}},\ \bibinfo {pages} {106802}
  (\bibinfo {year} {2011})}\BibitemShut {NoStop}%
\bibitem [{\citenamefont {Xu}\ \emph {et~al.}(2012)\citenamefont {Xu},
  \citenamefont {Liu}, \citenamefont {Alidoust}, \citenamefont {Neupane},
  \citenamefont {Qian}, \citenamefont {Belopolski}, \citenamefont {Denlinger},
  \citenamefont {Wang}, \citenamefont {Lin}, \citenamefont {Wray},
  \citenamefont {Landolt}, \citenamefont {Slomski}, \citenamefont {Dil},
  \citenamefont {Marcinkova}, \citenamefont {Morosan}, \citenamefont {Gibson},
  \citenamefont {Sankar}, \citenamefont {Chou}, \citenamefont {Cava},
  \citenamefont {Bansil},\ and\ \citenamefont {Hasan}}]{Xu12}%
  \BibitemOpen
  \bibfield  {author} {\bibinfo {author} {\bibfnamefont {S.-Y.}\ \bibnamefont
  {Xu}}, \bibinfo {author} {\bibfnamefont {C.}~\bibnamefont {Liu}}, \bibinfo
  {author} {\bibfnamefont {N.}~\bibnamefont {Alidoust}}, \bibinfo {author}
  {\bibfnamefont {M.}~\bibnamefont {Neupane}}, \bibinfo {author} {\bibfnamefont
  {D.}~\bibnamefont {Qian}}, \bibinfo {author} {\bibfnamefont {I.}~\bibnamefont
  {Belopolski}}, \bibinfo {author} {\bibfnamefont {J.~D.}\ \bibnamefont
  {Denlinger}}, \bibinfo {author} {\bibfnamefont {Y.~J.}\ \bibnamefont {Wang}},
  \bibinfo {author} {\bibfnamefont {H.}~\bibnamefont {Lin}}, \bibinfo {author}
  {\bibfnamefont {L.~A.}\ \bibnamefont {Wray}}, \bibinfo {author}
  {\bibfnamefont {G.}~\bibnamefont {Landolt}}, \bibinfo {author} {\bibfnamefont
  {B.}~\bibnamefont {Slomski}}, \bibinfo {author} {\bibfnamefont {J.~H.}\
  \bibnamefont {Dil}}, \bibinfo {author} {\bibfnamefont {A.}~\bibnamefont
  {Marcinkova}}, \bibinfo {author} {\bibfnamefont {E.}~\bibnamefont {Morosan}},
  \bibinfo {author} {\bibfnamefont {Q.}~\bibnamefont {Gibson}}, \bibinfo
  {author} {\bibfnamefont {R.}~\bibnamefont {Sankar}}, \bibinfo {author}
  {\bibfnamefont {F.~C.}\ \bibnamefont {Chou}}, \bibinfo {author}
  {\bibfnamefont {R.~J.}\ \bibnamefont {Cava}}, \bibinfo {author}
  {\bibfnamefont {A.}~\bibnamefont {Bansil}}, \ and\ \bibinfo {author}
  {\bibfnamefont {M.~Z.}\ \bibnamefont {Hasan}},\ }\href@noop {} {\bibfield
  {journal} {\bibinfo  {journal} {Nature}\ }\textbf {\bibinfo {volume} {3}},\
  \bibinfo {pages} {1192} (\bibinfo {year} {2012})}\BibitemShut {NoStop}%
\bibitem [{\citenamefont {Miyamoto}\ \emph
  {et~al.}(2012{\natexlab{a}})\citenamefont {Miyamoto}, \citenamefont {Kimura},
  \citenamefont {Kuroda}, \citenamefont {Okuda}, \citenamefont {Shimada},
  \citenamefont {Namatame}, \citenamefont {Taniguchi},\ and\ \citenamefont
  {Donath}}]{Miyamoto12}%
  \BibitemOpen
  \bibfield  {author} {\bibinfo {author} {\bibfnamefont {K.}~\bibnamefont
  {Miyamoto}}, \bibinfo {author} {\bibfnamefont {A.}~\bibnamefont {Kimura}},
  \bibinfo {author} {\bibfnamefont {K.}~\bibnamefont {Kuroda}}, \bibinfo
  {author} {\bibfnamefont {T.}~\bibnamefont {Okuda}}, \bibinfo {author}
  {\bibfnamefont {K.}~\bibnamefont {Shimada}}, \bibinfo {author} {\bibfnamefont
  {H.}~\bibnamefont {Namatame}}, \bibinfo {author} {\bibfnamefont
  {M.}~\bibnamefont {Taniguchi}}, \ and\ \bibinfo {author} {\bibfnamefont
  {M.}~\bibnamefont {Donath}},\ }\href@noop {} {\bibfield  {journal} {\bibinfo
  {journal} {Phys.\ Rev.\ Lett.}\ }\textbf {\bibinfo {volume} {108}},\ \bibinfo
  {pages} {066808} (\bibinfo {year} {2012}{\natexlab{a}})}\BibitemShut
  {NoStop}%
\bibitem [{\citenamefont {Miyamoto}\ \emph
  {et~al.}(2012{\natexlab{b}})\citenamefont {Miyamoto}, \citenamefont {Kimura},
  \citenamefont {Okuda}, \citenamefont {Shimada}, \citenamefont {Iwasawa},
  \citenamefont {Hayashi}, \citenamefont {Namatame}, \citenamefont
  {Taniguchi},\ and\ \citenamefont {Donath}}]{Miyamoto12c}%
  \BibitemOpen
  \bibfield  {author} {\bibinfo {author} {\bibfnamefont {K.}~\bibnamefont
  {Miyamoto}}, \bibinfo {author} {\bibfnamefont {A.}~\bibnamefont {Kimura}},
  \bibinfo {author} {\bibfnamefont {T.}~\bibnamefont {Okuda}}, \bibinfo
  {author} {\bibfnamefont {K.}~\bibnamefont {Shimada}}, \bibinfo {author}
  {\bibfnamefont {H.}~\bibnamefont {Iwasawa}}, \bibinfo {author} {\bibfnamefont
  {H.}~\bibnamefont {Hayashi}}, \bibinfo {author} {\bibfnamefont
  {H.}~\bibnamefont {Namatame}}, \bibinfo {author} {\bibfnamefont
  {M.}~\bibnamefont {Taniguchi}}, \ and\ \bibinfo {author} {\bibfnamefont
  {M.}~\bibnamefont {Donath}},\ }\href@noop {} {\bibfield  {journal} {\bibinfo
  {journal} {Phys.\ Rev.\ B}\ }\textbf {\bibinfo {volume} {86}},\ \bibinfo
  {pages} {161411(R)} (\bibinfo {year} {2012}{\natexlab{b}})}\BibitemShut
  {NoStop}%
\bibitem [{\citenamefont {Hochstrasser}\ \emph {et~al.}(2002)\citenamefont
  {Hochstrasser}, \citenamefont {Tobin}, \citenamefont {Rotenberg},\ and\
  \citenamefont {Kevan}}]{Hochstrasser02}%
  \BibitemOpen
  \bibfield  {author} {\bibinfo {author} {\bibfnamefont {M.}~\bibnamefont
  {Hochstrasser}}, \bibinfo {author} {\bibfnamefont {J.~G.}\ \bibnamefont
  {Tobin}}, \bibinfo {author} {\bibfnamefont {E.}~\bibnamefont {Rotenberg}}, \
  and\ \bibinfo {author} {\bibfnamefont {S.~D.}\ \bibnamefont {Kevan}},\
  }\href@noop {} {\bibfield  {journal} {\bibinfo  {journal} {Phys.\ Rev.\
  Lett.}\ }\textbf {\bibinfo {volume} {89}},\ \bibinfo {pages} {216802}
  (\bibinfo {year} {2002})}\BibitemShut {NoStop}%
\bibitem [{\citenamefont {Rybkin}\ \emph {et~al.}(2012)\citenamefont {Rybkin},
  \citenamefont {Krasovskii}, \citenamefont {Marchenko}, \citenamefont
  {Chulkov}, \citenamefont {Varykhalov}, \citenamefont {Rader},\ and\
  \citenamefont {Shikin}}]{Rybkin12}%
  \BibitemOpen
  \bibfield  {author} {\bibinfo {author} {\bibfnamefont {A.~G.}\ \bibnamefont
  {Rybkin}}, \bibinfo {author} {\bibfnamefont {E.~E.}\ \bibnamefont
  {Krasovskii}}, \bibinfo {author} {\bibfnamefont {D.}~\bibnamefont
  {Marchenko}}, \bibinfo {author} {\bibfnamefont {E.~V.}\ \bibnamefont
  {Chulkov}}, \bibinfo {author} {\bibfnamefont {A.}~\bibnamefont {Varykhalov}},
  \bibinfo {author} {\bibfnamefont {O.}~\bibnamefont {Rader}}, \ and\ \bibinfo
  {author} {\bibfnamefont {A.~M.}\ \bibnamefont {Shikin}},\ }\href@noop {}
  {\bibfield  {journal} {\bibinfo  {journal} {Phys.\ Rev.\ B}\ }\textbf
  {\bibinfo {volume} {86}},\ \bibinfo {pages} {035117} (\bibinfo {year}
  {2012})}\BibitemShut {NoStop}%
\bibitem [{\citenamefont {Giebels}\ \emph {et~al.}(2013)\citenamefont
  {Giebels}, \citenamefont {Gollisch},\ and\ \citenamefont
  {Feder}}]{Giebels13}%
  \BibitemOpen
  \bibfield  {author} {\bibinfo {author} {\bibfnamefont {F.}~\bibnamefont
  {Giebels}}, \bibinfo {author} {\bibfnamefont {H.}~\bibnamefont {Gollisch}}, \
  and\ \bibinfo {author} {\bibfnamefont {R.}~\bibnamefont {Feder}},\
  }\href@noop {} {\bibfield  {journal} {\bibinfo  {journal} {Phys.\ Rev.\ B}\
  }\textbf {\bibinfo {volume} {87}},\ \bibinfo {pages} {035124} (\bibinfo
  {year} {2013})}\BibitemShut {NoStop}%
\bibitem [{\citenamefont {Mirhosseini}\ \emph
  {et~al.}(2013{\natexlab{a}})\citenamefont {Mirhosseini}, \citenamefont
  {Flieger},\ and\ \citenamefont {Henk}}]{Mirhosseini13}%
  \BibitemOpen
  \bibfield  {author} {\bibinfo {author} {\bibfnamefont {H.}~\bibnamefont
  {Mirhosseini}}, \bibinfo {author} {\bibfnamefont {M.}~\bibnamefont
  {Flieger}}, \ and\ \bibinfo {author} {\bibfnamefont {J.}~\bibnamefont
  {Henk}},\ }\href@noop {} {\bibfield  {journal} {\bibinfo  {journal} {New J.
  Phys.}\ }\textbf {\bibinfo {volume} {15}},\ \bibinfo {pages} {033019}
  (\bibinfo {year} {2013}{\natexlab{a}})}\BibitemShut {NoStop}%
\bibitem [{\citenamefont {Mirhosseini}\ \emph
  {et~al.}(2013{\natexlab{b}})\citenamefont {Mirhosseini}, \citenamefont
  {Giebels}, \citenamefont {Gollisch}, \citenamefont {Henk},\ and\
  \citenamefont {Feder}}]{Mirhosseini13b}%
  \BibitemOpen
  \bibfield  {author} {\bibinfo {author} {\bibfnamefont {H.}~\bibnamefont
  {Mirhosseini}}, \bibinfo {author} {\bibfnamefont {F.}~\bibnamefont
  {Giebels}}, \bibinfo {author} {\bibfnamefont {H.}~\bibnamefont {Gollisch}},
  \bibinfo {author} {\bibfnamefont {J.}~\bibnamefont {Henk}}, \ and\ \bibinfo
  {author} {\bibfnamefont {R.}~\bibnamefont {Feder}},\ }\href@noop {}
  {\bibfield  {journal} {\bibinfo  {journal} {New J. Phys.}\ }\textbf {\bibinfo
  {volume} {15}},\ \bibinfo {pages} {095017} (\bibinfo {year}
  {2013}{\natexlab{b}})}\BibitemShut {NoStop}%
\bibitem [{\citenamefont {Braun}\ \emph {et~al.}(2014)\citenamefont {Braun},
  \citenamefont {Miyamoto}, \citenamefont {Kimura}, \citenamefont {Okuda},
  \citenamefont {Donath}, \citenamefont {Ebert},\ and\ \citenamefont
  {Min\'ar}}]{Braun14}%
  \BibitemOpen
  \bibfield  {author} {\bibinfo {author} {\bibfnamefont {J.}~\bibnamefont
  {Braun}}, \bibinfo {author} {\bibfnamefont {K.}~\bibnamefont {Miyamoto}},
  \bibinfo {author} {\bibfnamefont {A.}~\bibnamefont {Kimura}}, \bibinfo
  {author} {\bibfnamefont {T.}~\bibnamefont {Okuda}}, \bibinfo {author}
  {\bibfnamefont {M.}~\bibnamefont {Donath}}, \bibinfo {author} {\bibfnamefont
  {H.}~\bibnamefont {Ebert}}, \ and\ \bibinfo {author} {\bibfnamefont
  {J.}~\bibnamefont {Min\'ar}},\ }\href@noop {} {\bibfield  {journal} {\bibinfo
   {journal} {New J. Phys.}\ }\textbf {\bibinfo {volume} {16}},\ \bibinfo
  {pages} {015005} (\bibinfo {year} {2014})}\BibitemShut {NoStop}%
\bibitem [{\citenamefont {Bychkov}\ and\ \citenamefont
  {Rashba}(1984{\natexlab{a}})}]{Bychkov84b}%
  \BibitemOpen
  \bibfield  {author} {\bibinfo {author} {\bibfnamefont {Y.~A.}\ \bibnamefont
  {Bychkov}}\ and\ \bibinfo {author} {\bibfnamefont {E.~I.}\ \bibnamefont
  {Rashba}},\ }\href@noop {} {\bibfield  {journal} {\bibinfo  {journal} {J.
  Phys.\ C: Sol.\ State Phys.}\ }\textbf {\bibinfo {volume} {17}},\ \bibinfo
  {pages} {6039} (\bibinfo {year} {1984}{\natexlab{a}})}\BibitemShut {NoStop}%
\bibitem [{\citenamefont {Bychkov}\ and\ \citenamefont
  {Rashba}(1984{\natexlab{b}})}]{Bychkov84a}%
  \BibitemOpen
  \bibfield  {author} {\bibinfo {author} {\bibfnamefont {Y.~A.}\ \bibnamefont
  {Bychkov}}\ and\ \bibinfo {author} {\bibfnamefont {E.~I.}\ \bibnamefont
  {Rashba}},\ }\href@noop {} {\bibfield  {journal} {\bibinfo  {journal} {Sov.\
  Phys.\ JETP Lett.}\ }\textbf {\bibinfo {volume} {39}},\ \bibinfo {pages} {78}
  (\bibinfo {year} {1984}{\natexlab{b}})},\ \bibinfo {note} {translated from
  Ref.~\cite{Bychkov84c}}\BibitemShut {NoStop}%
\bibitem [{\citenamefont {Winkler}(2003)}]{Winkler03}%
  \BibitemOpen
  \bibfield  {author} {\bibinfo {author} {\bibfnamefont {R.}~\bibnamefont
  {Winkler}},\ }\href@noop {} {\emph {\bibinfo {title} {Spin-Orbit Coupling
  Effects in Two-Dimensional Electron and Hole Systems}}}\ (\bibinfo
  {publisher} {Springer},\ \bibinfo {address} {Berlin},\ \bibinfo {year}
  {2003})\BibitemShut {NoStop}%
\bibitem [{\citenamefont {Eremeev}\ \emph {et~al.}(2012)\citenamefont
  {Eremeev}, \citenamefont {Landolt}, \citenamefont {Menshchikova},
  \citenamefont {Slomski}, \citenamefont {Koroteev}, \citenamefont {Aliev},
  \citenamefont {Babanly}, \citenamefont {Henk}, \citenamefont {Ernst},
  \citenamefont {Patthey}, \citenamefont {Eich}, \citenamefont {Khajetoorians},
  \citenamefont {Hagemeister}, \citenamefont {Pietzsch}, \citenamefont {Wiebe},
  \citenamefont {Wiesendanger}, \citenamefont {Echenique}, \citenamefont
  {Tsirkin}, \citenamefont {Amiraslanov}, \citenamefont {Dil},\ and\
  \citenamefont {Chulkov}}]{Eremeev12}%
  \BibitemOpen
  \bibfield  {author} {\bibinfo {author} {\bibfnamefont {S.~V.}\ \bibnamefont
  {Eremeev}}, \bibinfo {author} {\bibfnamefont {G.}~\bibnamefont {Landolt}},
  \bibinfo {author} {\bibfnamefont {T.~V.}\ \bibnamefont {Menshchikova}},
  \bibinfo {author} {\bibfnamefont {B.}~\bibnamefont {Slomski}}, \bibinfo
  {author} {\bibfnamefont {Y.~M.}\ \bibnamefont {Koroteev}}, \bibinfo {author}
  {\bibfnamefont {Z.~S.}\ \bibnamefont {Aliev}}, \bibinfo {author}
  {\bibfnamefont {M.~B.}\ \bibnamefont {Babanly}}, \bibinfo {author}
  {\bibfnamefont {J.}~\bibnamefont {Henk}}, \bibinfo {author} {\bibfnamefont
  {A.}~\bibnamefont {Ernst}}, \bibinfo {author} {\bibfnamefont
  {L.}~\bibnamefont {Patthey}}, \bibinfo {author} {\bibfnamefont
  {A.}~\bibnamefont {Eich}}, \bibinfo {author} {\bibfnamefont {A.~A.}\
  \bibnamefont {Khajetoorians}}, \bibinfo {author} {\bibfnamefont
  {J.}~\bibnamefont {Hagemeister}}, \bibinfo {author} {\bibfnamefont
  {O.}~\bibnamefont {Pietzsch}}, \bibinfo {author} {\bibfnamefont
  {J.}~\bibnamefont {Wiebe}}, \bibinfo {author} {\bibfnamefont
  {R.}~\bibnamefont {Wiesendanger}}, \bibinfo {author} {\bibfnamefont {P.~M.}\
  \bibnamefont {Echenique}}, \bibinfo {author} {\bibfnamefont {S.~S.}\
  \bibnamefont {Tsirkin}}, \bibinfo {author} {\bibfnamefont {I.~R.}\
  \bibnamefont {Amiraslanov}}, \bibinfo {author} {\bibfnamefont {J.~H.}\
  \bibnamefont {Dil}}, \ and\ \bibinfo {author} {\bibfnamefont {E.~V.}\
  \bibnamefont {Chulkov}},\ }\href@noop {} {\bibfield  {journal} {\bibinfo
  {journal} {Nature Comms.}\ }\textbf {\bibinfo {volume} {3}},\ \bibinfo
  {pages} {635} (\bibinfo {year} {2012})}\BibitemShut {NoStop}%
\bibitem [{\citenamefont {Bianchi}\ \emph {et~al.}(2010)\citenamefont
  {Bianchi}, \citenamefont {Guan}, \citenamefont {Bao}, \citenamefont {Mi},
  \citenamefont {Iversen}, \citenamefont {King},\ and\ \citenamefont
  {Hofmann}}]{Bianchi10}%
  \BibitemOpen
  \bibfield  {author} {\bibinfo {author} {\bibfnamefont {M.}~\bibnamefont
  {Bianchi}}, \bibinfo {author} {\bibfnamefont {D.}~\bibnamefont {Guan}},
  \bibinfo {author} {\bibfnamefont {S.}~\bibnamefont {Bao}}, \bibinfo {author}
  {\bibfnamefont {J.}~\bibnamefont {Mi}}, \bibinfo {author} {\bibfnamefont
  {B.~B.}\ \bibnamefont {Iversen}}, \bibinfo {author} {\bibfnamefont
  {P.~D.~C.}\ \bibnamefont {King}}, \ and\ \bibinfo {author} {\bibfnamefont
  {P.}~\bibnamefont {Hofmann}},\ }\href@noop {} {\bibfield  {journal} {\bibinfo
   {journal} {Nature Comms.}\ }\textbf {\bibinfo {volume} {1}},\ \bibinfo
  {pages} {128} (\bibinfo {year} {2010})}\BibitemShut {NoStop}%
\bibitem [{\citenamefont {Yang}\ \emph {et~al.}(2015)\citenamefont {Yang},
  \citenamefont {Liu}, \citenamefont {Sun}, \citenamefont {Peng}, \citenamefont
  {Yang}, \citenamefont {Zhang}, \citenamefont {Zhou}, \citenamefont {Zhang},
  \citenamefont {Guo}, \citenamefont {Rahn}, \citenamefont {Prabhakaran},
  \citenamefont {Hussain}, \citenamefont {Mo}, \citenamefont {Felser},
  \citenamefont {Yan},\ and\ \citenamefont {Chen}}]{Yang15}%
  \BibitemOpen
  \bibfield  {author} {\bibinfo {author} {\bibfnamefont {L.~X.}\ \bibnamefont
  {Yang}}, \bibinfo {author} {\bibfnamefont {Z.~K.}\ \bibnamefont {Liu}},
  \bibinfo {author} {\bibfnamefont {Y.}~\bibnamefont {Sun}}, \bibinfo {author}
  {\bibfnamefont {H.}~\bibnamefont {Peng}}, \bibinfo {author} {\bibfnamefont
  {H.~F.}\ \bibnamefont {Yang}}, \bibinfo {author} {\bibfnamefont
  {T.}~\bibnamefont {Zhang}}, \bibinfo {author} {\bibfnamefont
  {B.}~\bibnamefont {Zhou}}, \bibinfo {author} {\bibfnamefont {Y.}~\bibnamefont
  {Zhang}}, \bibinfo {author} {\bibfnamefont {Y.~F.}\ \bibnamefont {Guo}},
  \bibinfo {author} {\bibfnamefont {M.}~\bibnamefont {Rahn}}, \bibinfo {author}
  {\bibfnamefont {D.}~\bibnamefont {Prabhakaran}}, \bibinfo {author}
  {\bibfnamefont {Z.}~\bibnamefont {Hussain}}, \bibinfo {author} {\bibfnamefont
  {S.~K.}\ \bibnamefont {Mo}}, \bibinfo {author} {\bibfnamefont
  {C.}~\bibnamefont {Felser}}, \bibinfo {author} {\bibfnamefont
  {B.}~\bibnamefont {Yan}}, \ and\ \bibinfo {author} {\bibfnamefont {Y.~L.}\
  \bibnamefont {Chen}},\ }\href@noop {} {\bibfield  {journal} {\bibinfo
  {journal} {Nature Physics}\ }\textbf {\bibinfo {volume} {11}},\ \bibinfo
  {pages} {728} (\bibinfo {year} {2015})}\BibitemShut {NoStop}%
\bibitem [{\citenamefont {Wang}\ \emph {et~al.}(2015)\citenamefont {Wang},
  \citenamefont {Deorani}, \citenamefont {Banerjee}, \citenamefont {Koirala},
  \citenamefont {Brahlek}, \citenamefont {Oh},\ and\ \citenamefont
  {Yang}}]{Wang15}%
  \BibitemOpen
  \bibfield  {author} {\bibinfo {author} {\bibfnamefont {Y.}~\bibnamefont
  {Wang}}, \bibinfo {author} {\bibfnamefont {P.}~\bibnamefont {Deorani}},
  \bibinfo {author} {\bibfnamefont {K.}~\bibnamefont {Banerjee}}, \bibinfo
  {author} {\bibfnamefont {N.}~\bibnamefont {Koirala}}, \bibinfo {author}
  {\bibfnamefont {M.}~\bibnamefont {Brahlek}}, \bibinfo {author} {\bibfnamefont
  {S.}~\bibnamefont {Oh}}, \ and\ \bibinfo {author} {\bibfnamefont
  {H.}~\bibnamefont {Yang}},\ }\href@noop {} {\bibfield  {journal} {\bibinfo
  {journal} {Phys. Rev. Lett.}\ }\textbf {\bibinfo {volume} {114}},\ \bibinfo
  {pages} {257202} (\bibinfo {year} {2015})}\BibitemShut {NoStop}%
\bibitem [{\citenamefont {Sun}\ \emph {et~al.}(2025)\citenamefont {Sun},
  \citenamefont {Wu},\ and\ \citenamefont {Yan}}]{Sun15}%
  \BibitemOpen
  \bibfield  {author} {\bibinfo {author} {\bibfnamefont {Y.}~\bibnamefont
  {Sun}}, \bibinfo {author} {\bibfnamefont {S.-C.}\ \bibnamefont {Wu}}, \ and\
  \bibinfo {author} {\bibfnamefont {B.}~\bibnamefont {Yan}},\ }\href@noop {}
  {\bibfield  {journal} {\bibinfo  {journal} {PRB}\ }\textbf {\bibinfo {volume}
  {92}},\ \bibinfo {pages} {115428} (\bibinfo {year} {2025})}\BibitemShut
  {NoStop}%
\bibitem [{\citenamefont {Yan}\ \emph {et~al.}(2013)\citenamefont {Yan},
  \citenamefont {Jansen},\ and\ \citenamefont {Felser}}]{Yan13}%
  \BibitemOpen
  \bibfield  {author} {\bibinfo {author} {\bibfnamefont {B.}~\bibnamefont
  {Yan}}, \bibinfo {author} {\bibfnamefont {M.}~\bibnamefont {Jansen}}, \ and\
  \bibinfo {author} {\bibfnamefont {C.}~\bibnamefont {Felser}},\ }\href@noop {}
  {\bibfield  {journal} {\bibinfo  {journal} {Nature Physics}\ }\textbf
  {\bibinfo {volume} {9}},\ \bibinfo {pages} {709} (\bibinfo {year}
  {2013})}\BibitemShut {NoStop}%
\bibitem [{\citenamefont {Li}\ \emph {et~al.}(2015)\citenamefont {Li},
  \citenamefont {Yan}, \citenamefont {Thomale},\ and\ \citenamefont
  {Hanke}}]{Li15}%
  \BibitemOpen
  \bibfield  {author} {\bibinfo {author} {\bibfnamefont {G.}~\bibnamefont
  {Li}}, \bibinfo {author} {\bibfnamefont {B.}~\bibnamefont {Yan}}, \bibinfo
  {author} {\bibfnamefont {R.}~\bibnamefont {Thomale}}, \ and\ \bibinfo
  {author} {\bibfnamefont {W.}~\bibnamefont {Hanke}},\ }\href@noop {}
  {\bibfield  {journal} {\bibinfo  {journal} {Scientific Reports}\ }\textbf
  {\bibinfo {volume} {5}},\ \bibinfo {pages} {10435} (\bibinfo {year}
  {2015})}\BibitemShut {NoStop}%
\bibitem [{\citenamefont {Yan}\ \emph {et~al.}(2015)\citenamefont {Yan},
  \citenamefont {Stadtm{\"u}ller}, \citenamefont {Haag}, \citenamefont
  {Jakobs}, \citenamefont {Seidel}, \citenamefont {Jungkenn}, \citenamefont
  {Mathias}, \citenamefont {Cinchetti}, \citenamefont {Aeschlimann},\ and\
  \citenamefont {Felser}}]{Yan15}%
  \BibitemOpen
  \bibfield  {author} {\bibinfo {author} {\bibfnamefont {B.}~\bibnamefont
  {Yan}}, \bibinfo {author} {\bibfnamefont {B.}~\bibnamefont
  {Stadtm{\"u}ller}}, \bibinfo {author} {\bibfnamefont {N.}~\bibnamefont
  {Haag}}, \bibinfo {author} {\bibfnamefont {S.}~\bibnamefont {Jakobs}},
  \bibinfo {author} {\bibfnamefont {J.}~\bibnamefont {Seidel}}, \bibinfo
  {author} {\bibfnamefont {D.}~\bibnamefont {Jungkenn}}, \bibinfo {author}
  {\bibfnamefont {S.}~\bibnamefont {Mathias}}, \bibinfo {author} {\bibfnamefont
  {M.}~\bibnamefont {Cinchetti}}, \bibinfo {author} {\bibfnamefont
  {M.}~\bibnamefont {Aeschlimann}}, \ and\ \bibinfo {author} {\bibfnamefont
  {C.}~\bibnamefont {Felser}},\ }\href@noop {} {\bibfield  {journal} {\bibinfo
  {journal} {Nature Comms.}\ }\textbf {\bibinfo {volume} {6}},\ \bibinfo
  {pages} {10167} (\bibinfo {year} {2015})}\BibitemShut {NoStop}%
\bibitem [{\citenamefont {Perdew}\ \emph {et~al.}(1996)\citenamefont {Perdew},
  \citenamefont {Burke},\ and\ \citenamefont {Ernzerhof}}]{Perdew96}%
  \BibitemOpen
  \bibfield  {author} {\bibinfo {author} {\bibfnamefont {J.~P.}\ \bibnamefont
  {Perdew}}, \bibinfo {author} {\bibfnamefont {K.}~\bibnamefont {Burke}}, \
  and\ \bibinfo {author} {\bibfnamefont {M.}~\bibnamefont {Ernzerhof}},\
  }\href@noop {} {\bibfield  {journal} {\bibinfo  {journal} {Phys.\ Rev.\
  Lett.}\ }\textbf {\bibinfo {volume} {77}},\ \bibinfo {pages} {3865} (\bibinfo
  {year} {1996})}\BibitemShut {NoStop}%
\bibitem [{\citenamefont {Perdew}\ \emph {et~al.}(1997)\citenamefont {Perdew},
  \citenamefont {Burke},\ and\ \citenamefont {Ernzerhof}}]{Perdew97}%
  \BibitemOpen
  \bibfield  {author} {\bibinfo {author} {\bibfnamefont {P.}~\bibnamefont
  {Perdew}}, \bibinfo {author} {\bibfnamefont {K.}~\bibnamefont {Burke}}, \
  and\ \bibinfo {author} {\bibfnamefont {M.}~\bibnamefont {Ernzerhof}},\
  }\href@noop {} {\bibfield  {journal} {\bibinfo  {journal} {Phys.\ Rev.\
  Lett.}\ }\textbf {\bibinfo {volume} {78}},\ \bibinfo {pages} {1396} (\bibinfo
  {year} {1997})}\BibitemShut {NoStop}%
\bibitem [{Note1()}]{Note1}%
  \BibitemOpen
  \bibinfo {note} {For details, see the online Supplementary
  Material.}\BibitemShut {Stop}%
\bibitem [{\citenamefont {Henk}(2002)}]{Henk02c}%
  \BibitemOpen
  \bibfield  {author} {\bibinfo {author} {\bibfnamefont {J.}~\bibnamefont
  {Henk}},\ }in\ \href@noop {} {\emph {\bibinfo {booktitle} {Handbook of Thin
  Film Materials}}},\ Vol.~\bibinfo {volume} {2},\ \bibinfo {editor} {edited
  by\ \bibinfo {editor} {\bibfnamefont {H.~S.}\ \bibnamefont {Nalwa}}}\
  (\bibinfo  {publisher} {Academic Press},\ \bibinfo {address} {San Diego},\
  \bibinfo {year} {2002})\ Chap.~\bibinfo {chapter} {10}, p.\ \bibinfo {pages}
  {479}\BibitemShut {NoStop}%
\bibitem [{\citenamefont {Zabloudil}\ \emph {et~al.}(2005)\citenamefont
  {Zabloudil}, \citenamefont {Hammerling}, \citenamefont {Szunyogh},\ and\
  \citenamefont {Weinberger}}]{Zabloudil05}%
  \BibitemOpen
  \bibinfo {editor} {\bibfnamefont {J.}~\bibnamefont {Zabloudil}}, \bibinfo
  {editor} {\bibfnamefont {R.}~\bibnamefont {Hammerling}}, \bibinfo {editor}
  {\bibfnamefont {L.}~\bibnamefont {Szunyogh}}, \ and\ \bibinfo {editor}
  {\bibfnamefont {P.}~\bibnamefont {Weinberger}},\ eds.,\ \href@noop {} {\emph
  {\bibinfo {title} {Electron Scattering in Solid Matter}}}\ (\bibinfo
  {publisher} {Springer},\ \bibinfo {address} {Berlin},\ \bibinfo {year}
  {2005})\BibitemShut {NoStop}%
\bibitem [{\citenamefont {Kresse}\ and\ \citenamefont
  {Furthm\"uller}(1996{\natexlab{a}})}]{Kresse96b}%
  \BibitemOpen
  \bibfield  {author} {\bibinfo {author} {\bibfnamefont {G.}~\bibnamefont
  {Kresse}}\ and\ \bibinfo {author} {\bibfnamefont {J.}~\bibnamefont
  {Furthm\"uller}},\ }\href@noop {} {\bibfield  {journal} {\bibinfo  {journal}
  {Comp. Mater. Sci.}\ }\textbf {\bibinfo {volume} {6}},\ \bibinfo {pages} {15}
  (\bibinfo {year} {1996}{\natexlab{a}})}\BibitemShut {NoStop}%
\bibitem [{\citenamefont {Kresse}\ and\ \citenamefont
  {Furthm\"uller}(1996{\natexlab{b}})}]{Kresse96a}%
  \BibitemOpen
  \bibfield  {author} {\bibinfo {author} {\bibfnamefont {G.}~\bibnamefont
  {Kresse}}\ and\ \bibinfo {author} {\bibfnamefont {J.}~\bibnamefont
  {Furthm\"uller}},\ }\href@noop {} {\bibfield  {journal} {\bibinfo  {journal}
  {Phys.\ Rev.\ B}\ }\textbf {\bibinfo {volume} {54}},\ \bibinfo {pages}
  {11169} (\bibinfo {year} {1996}{\natexlab{b}})}\BibitemShut {NoStop}%
\bibitem [{\citenamefont {Venus}\ \emph {et~al.}(2000)\citenamefont {Venus},
  \citenamefont {Cool},\ and\ \citenamefont {Plihal}}]{Venus00}%
  \BibitemOpen
  \bibfield  {author} {\bibinfo {author} {\bibfnamefont {D.}~\bibnamefont
  {Venus}}, \bibinfo {author} {\bibfnamefont {S.}~\bibnamefont {Cool}}, \ and\
  \bibinfo {author} {\bibfnamefont {M.}~\bibnamefont {Plihal}},\ }\href@noop {}
  {\bibfield  {journal} {\bibinfo  {journal} {Surf.\ Sci.}\ }\textbf {\bibinfo
  {volume} {446}},\ \bibinfo {pages} {199} (\bibinfo {year}
  {2000})}\BibitemShut {NoStop}%
\bibitem [{\citenamefont {Meyerheim}\ \emph {et~al.}(2001)\citenamefont
  {Meyerheim}, \citenamefont {Sander}, \citenamefont {Popescu}, \citenamefont
  {Steadman}, \citenamefont {Ferrer},\ and\ \citenamefont
  {Kirschner}}]{Meyerheim01a}%
  \BibitemOpen
  \bibfield  {author} {\bibinfo {author} {\bibfnamefont {H.~L.}\ \bibnamefont
  {Meyerheim}}, \bibinfo {author} {\bibfnamefont {D.}~\bibnamefont {Sander}},
  \bibinfo {author} {\bibfnamefont {R.}~\bibnamefont {Popescu}}, \bibinfo
  {author} {\bibfnamefont {P.}~\bibnamefont {Steadman}}, \bibinfo {author}
  {\bibfnamefont {S.}~\bibnamefont {Ferrer}}, \ and\ \bibinfo {author}
  {\bibfnamefont {J.}~\bibnamefont {Kirschner}},\ }\href@noop {} {\bibfield
  {journal} {\bibinfo  {journal} {Surf.\ Sci.}\ }\textbf {\bibinfo {volume}
  {475}},\ \bibinfo {pages} {103} (\bibinfo {year} {2001})}\BibitemShut
  {NoStop}%
\bibitem [{\citenamefont {Engelkamp}\ \emph {et~al.}(2015)\citenamefont
  {Engelkamp}, \citenamefont {Wortelen}, \citenamefont {Mirhosseini},
  \citenamefont {Schmidt}, \citenamefont {Thonig}, \citenamefont {Henk},\ and\
  \citenamefont {Donath}}]{Engelkamp15}%
  \BibitemOpen
  \bibfield  {author} {\bibinfo {author} {\bibfnamefont {B.}~\bibnamefont
  {Engelkamp}}, \bibinfo {author} {\bibfnamefont {H.}~\bibnamefont {Wortelen}},
  \bibinfo {author} {\bibfnamefont {H.}~\bibnamefont {Mirhosseini}}, \bibinfo
  {author} {\bibfnamefont {A.~B.}\ \bibnamefont {Schmidt}}, \bibinfo {author}
  {\bibfnamefont {D.}~\bibnamefont {Thonig}}, \bibinfo {author} {\bibfnamefont
  {J.}~\bibnamefont {Henk}}, \ and\ \bibinfo {author} {\bibfnamefont
  {M.}~\bibnamefont {Donath}},\ }\href@noop {} {\bibfield  {journal} {\bibinfo
  {journal} {Phys.\ Rev.\ B}\ }\textbf {\bibinfo {volume} {92}},\ \bibinfo
  {pages} {085401} (\bibinfo {year} {2015})}\BibitemShut {NoStop}%
\bibitem [{\citenamefont {Wortelen}\ \emph {et~al.}(2015)\citenamefont
  {Wortelen}, \citenamefont {Miyamoto}, \citenamefont {Mirhosseini},
  \citenamefont {Okuda}, \citenamefont {Kimura}, \citenamefont {Thonig},
  \citenamefont {Henk},\ and\ \citenamefont {Donath}}]{Wortelen15b}%
  \BibitemOpen
  \bibfield  {author} {\bibinfo {author} {\bibfnamefont {H.}~\bibnamefont
  {Wortelen}}, \bibinfo {author} {\bibfnamefont {K.}~\bibnamefont {Miyamoto}},
  \bibinfo {author} {\bibfnamefont {H.}~\bibnamefont {Mirhosseini}}, \bibinfo
  {author} {\bibfnamefont {T.}~\bibnamefont {Okuda}}, \bibinfo {author}
  {\bibfnamefont {A.}~\bibnamefont {Kimura}}, \bibinfo {author} {\bibfnamefont
  {D.}~\bibnamefont {Thonig}}, \bibinfo {author} {\bibfnamefont
  {J.}~\bibnamefont {Henk}}, \ and\ \bibinfo {author} {\bibfnamefont
  {M.}~\bibnamefont {Donath}},\ }\href@noop {} {\bibfield  {journal} {\bibinfo
  {journal} {Phys.\ Rev.\ B}\ }\textbf {\bibinfo {volume} {92}},\ \bibinfo
  {pages} {161408(R)} (\bibinfo {year} {2015})}\BibitemShut {NoStop}%
\bibitem [{\citenamefont {Rauch}\ \emph {et~al.}(2014)\citenamefont {Rauch},
  \citenamefont {Flieger}, \citenamefont {Henk}, \citenamefont {Mertig},\ and\
  \citenamefont {Ernst}}]{Rauch14}%
  \BibitemOpen
  \bibfield  {author} {\bibinfo {author} {\bibfnamefont {T.}~\bibnamefont
  {Rauch}}, \bibinfo {author} {\bibfnamefont {M.}~\bibnamefont {Flieger}},
  \bibinfo {author} {\bibfnamefont {J.}~\bibnamefont {Henk}}, \bibinfo {author}
  {\bibfnamefont {I.}~\bibnamefont {Mertig}}, \ and\ \bibinfo {author}
  {\bibfnamefont {A.}~\bibnamefont {Ernst}},\ }\href@noop {} {\bibfield
  {journal} {\bibinfo  {journal} {Phys.\ Rev.\ Lett.}\ }\textbf {\bibinfo
  {volume} {112}},\ \bibinfo {pages} {016802} (\bibinfo {year}
  {2014})}\BibitemShut {NoStop}%
\bibitem [{\citenamefont {Barone}\ \emph {et~al.}(2013)\citenamefont {Barone},
  \citenamefont {Rauch}, \citenamefont {Di~Sante}, \citenamefont {Henk},
  \citenamefont {Mertig},\ and\ \citenamefont {Picozzi}}]{Barone13}%
  \BibitemOpen
  \bibfield  {author} {\bibinfo {author} {\bibfnamefont {P.}~\bibnamefont
  {Barone}}, \bibinfo {author} {\bibfnamefont {T.}~\bibnamefont {Rauch}},
  \bibinfo {author} {\bibfnamefont {D.}~\bibnamefont {Di~Sante}}, \bibinfo
  {author} {\bibfnamefont {J.}~\bibnamefont {Henk}}, \bibinfo {author}
  {\bibfnamefont {I.}~\bibnamefont {Mertig}}, \ and\ \bibinfo {author}
  {\bibfnamefont {S.}~\bibnamefont {Picozzi}},\ }\href@noop {} {\bibfield
  {journal} {\bibinfo  {journal} {Phys.\ Rev.\ B}\ }\textbf {\bibinfo {volume}
  {88}},\ \bibinfo {pages} {045207} (\bibinfo {year} {2013})}\BibitemShut
  {NoStop}%
\bibitem [{\citenamefont {Rauch}\ \emph {et~al.}(2015)\citenamefont {Rauch},
  \citenamefont {Achilles}, \citenamefont {Henk},\ and\ \citenamefont
  {Mertig}}]{Rauch15}%
  \BibitemOpen
  \bibfield  {author} {\bibinfo {author} {\bibfnamefont {T.}~\bibnamefont
  {Rauch}}, \bibinfo {author} {\bibfnamefont {S.}~\bibnamefont {Achilles}},
  \bibinfo {author} {\bibfnamefont {J.}~\bibnamefont {Henk}}, \ and\ \bibinfo
  {author} {\bibfnamefont {I.}~\bibnamefont {Mertig}},\ }\href@noop {}
  {\bibfield  {journal} {\bibinfo  {journal} {Phys.\ Rev.\ Lett.}\ }\textbf
  {\bibinfo {volume} {114}},\ \bibinfo {pages} {236805} (\bibinfo {year}
  {2015})}\BibitemShut {NoStop}%
\bibitem [{\citenamefont {Miyamoto}\ \emph {et~al.}(2015)\citenamefont
  {Miyamoto}, \citenamefont {Kimura}, \citenamefont {Okuda},\ and\
  \citenamefont {Donath}}]{Miyamoto15}%
  \BibitemOpen
  \bibfield  {author} {\bibinfo {author} {\bibfnamefont {K.}~\bibnamefont
  {Miyamoto}}, \bibinfo {author} {\bibfnamefont {A.}~\bibnamefont {Kimura}},
  \bibinfo {author} {\bibfnamefont {T.}~\bibnamefont {Okuda}}, \ and\ \bibinfo
  {author} {\bibfnamefont {M.}~\bibnamefont {Donath}},\ }\href@noop {}
  {\bibfield  {journal} {\bibinfo  {journal} {Journal of Electron Spectroscopy
  and Related Phenomena}\ }\textbf {\bibinfo {volume} {201}},\ \bibinfo {pages}
  {53} (\bibinfo {year} {2015})}\BibitemShut {NoStop}%
\bibitem [{\citenamefont {Miyamoto}\ \emph {et~al.}(2016)\citenamefont
  {Miyamoto}, \citenamefont {Wortelen}, \citenamefont {Mirhosseini},
  \citenamefont {Okuda}, \citenamefont {Kimura}, \citenamefont {Iwasawa},
  \citenamefont {Shimada}, \citenamefont {Henk},\ and\ \citenamefont
  {Donath}}]{Miyamoto16}%
  \BibitemOpen
  \bibfield  {author} {\bibinfo {author} {\bibfnamefont {K.}~\bibnamefont
  {Miyamoto}}, \bibinfo {author} {\bibfnamefont {H.}~\bibnamefont {Wortelen}},
  \bibinfo {author} {\bibfnamefont {H.}~\bibnamefont {Mirhosseini}}, \bibinfo
  {author} {\bibfnamefont {T.}~\bibnamefont {Okuda}}, \bibinfo {author}
  {\bibfnamefont {A.}~\bibnamefont {Kimura}}, \bibinfo {author} {\bibfnamefont
  {H.}~\bibnamefont {Iwasawa}}, \bibinfo {author} {\bibfnamefont
  {K.}~\bibnamefont {Shimada}}, \bibinfo {author} {\bibfnamefont
  {J.}~\bibnamefont {Henk}}, \ and\ \bibinfo {author} {\bibfnamefont
  {M.}~\bibnamefont {Donath}},\ }\href@noop {} {\bibfield  {journal} {\bibinfo
  {journal} {Phys.\ Rev.\ B}\ }\textbf {\bibinfo {volume} {93}},\ \bibinfo
  {pages} {161403(R)} (\bibinfo {year} {2016})}\BibitemShut {NoStop}%
\bibitem [{\citenamefont {Stolwijk}\ \emph {et~al.}(2014)\citenamefont
  {Stolwijk}, \citenamefont {Wortelen}, \citenamefont {Schmidt},\ and\
  \citenamefont {Donath}}]{Stolwijk14}%
  \BibitemOpen
  \bibfield  {author} {\bibinfo {author} {\bibfnamefont {S.~D.}\ \bibnamefont
  {Stolwijk}}, \bibinfo {author} {\bibfnamefont {H.}~\bibnamefont {Wortelen}},
  \bibinfo {author} {\bibfnamefont {A.~B.}\ \bibnamefont {Schmidt}}, \ and\
  \bibinfo {author} {\bibfnamefont {M.}~\bibnamefont {Donath}},\ }\href@noop {}
  {\bibfield  {journal} {\bibinfo  {journal} {Rev.\ Sci.\ Instrum.}\ }\textbf
  {\bibinfo {volume} {85}},\ \bibinfo {pages} {013306} (\bibinfo {year}
  {2014})}\BibitemShut {NoStop}%
\bibitem [{\citenamefont {Budke}\ \emph {et~al.}(2007)\citenamefont {Budke},
  \citenamefont {Allmers}, \citenamefont {Donath},\ and\ \citenamefont
  {Rangelov}}]{Budke07}%
  \BibitemOpen
  \bibfield  {author} {\bibinfo {author} {\bibfnamefont {M.}~\bibnamefont
  {Budke}}, \bibinfo {author} {\bibfnamefont {T.}~\bibnamefont {Allmers}},
  \bibinfo {author} {\bibfnamefont {M.}~\bibnamefont {Donath}}, \ and\ \bibinfo
  {author} {\bibfnamefont {G.}~\bibnamefont {Rangelov}},\ }\href@noop {}
  {\bibfield  {journal} {\bibinfo  {journal} {Rev.\ Sci.\ Instrum.}\ }\textbf
  {\bibinfo {volume} {78}},\ \bibinfo {pages} {113909} (\bibinfo {year}
  {2007})}\BibitemShut {NoStop}%
\bibitem [{Note2()}]{Note2}%
  \BibitemOpen
  \bibinfo {note} {A \protect \emph {global} gap separates two bands for all
  wave vectors $\protect \ensuremath {\protect \bm {k}}$; there is a $\protect
  \ensuremath {\protect \bm {k}}$-independent energy range which separates the
  upper from the lower band. In a \protect \emph {semimetal} gap there is a
  $\protect \ensuremath {\protect \bm {k}}$-dependent energy range separating
  the two bands.}\BibitemShut {Stop}%
\bibitem [{\citenamefont {LaShell}\ \emph {et~al.}(1996)\citenamefont
  {LaShell}, \citenamefont {McDougall},\ and\ \citenamefont
  {Jensen}}]{LaShell96}%
  \BibitemOpen
  \bibfield  {author} {\bibinfo {author} {\bibfnamefont {S.}~\bibnamefont
  {LaShell}}, \bibinfo {author} {\bibfnamefont {B.~A.}\ \bibnamefont
  {McDougall}}, \ and\ \bibinfo {author} {\bibfnamefont {E.}~\bibnamefont
  {Jensen}},\ }\href@noop {} {\bibfield  {journal} {\bibinfo  {journal} {Phys.\
  Rev.\ Lett.}\ }\textbf {\bibinfo {volume} {77}},\ \bibinfo {pages} {3419}
  (\bibinfo {year} {1996})}\BibitemShut {NoStop}%
\bibitem [{\citenamefont {Ast}\ \emph {et~al.}(2007)\citenamefont {Ast},
  \citenamefont {Henk}, \citenamefont {Ernst}, \citenamefont {Moreschini},
  \citenamefont {Falub}, \citenamefont {Pacil\'{e}}, \citenamefont {Bruno},
  \citenamefont {Kern},\ and\ \citenamefont {Grioni}}]{Ast07a}%
  \BibitemOpen
  \bibfield  {author} {\bibinfo {author} {\bibfnamefont {C.~R.}\ \bibnamefont
  {Ast}}, \bibinfo {author} {\bibfnamefont {J.}~\bibnamefont {Henk}}, \bibinfo
  {author} {\bibfnamefont {A.}~\bibnamefont {Ernst}}, \bibinfo {author}
  {\bibfnamefont {L.}~\bibnamefont {Moreschini}}, \bibinfo {author}
  {\bibfnamefont {M.~C.}\ \bibnamefont {Falub}}, \bibinfo {author}
  {\bibfnamefont {D.}~\bibnamefont {Pacil\'{e}}}, \bibinfo {author}
  {\bibfnamefont {P.}~\bibnamefont {Bruno}}, \bibinfo {author} {\bibfnamefont
  {K.}~\bibnamefont {Kern}}, \ and\ \bibinfo {author} {\bibfnamefont
  {M.}~\bibnamefont {Grioni}},\ }\href@noop {} {\bibfield  {journal} {\bibinfo
  {journal} {Phys.\ Rev.\ Lett.}\ }\textbf {\bibinfo {volume} {98}},\ \bibinfo
  {pages} {186807} (\bibinfo {year} {2007})}\BibitemShut {NoStop}%
\bibitem [{\citenamefont {Requist}\ \emph {et~al.}(2015)\citenamefont
  {Requist}, \citenamefont {Sheverdyaeva}, \citenamefont {Moras}, \citenamefont
  {Mahatha}, \citenamefont {Carbone},\ and\ \citenamefont
  {Tosatti}}]{Requist15}%
  \BibitemOpen
  \bibfield  {author} {\bibinfo {author} {\bibfnamefont {R.}~\bibnamefont
  {Requist}}, \bibinfo {author} {\bibfnamefont {P.~M.}\ \bibnamefont
  {Sheverdyaeva}}, \bibinfo {author} {\bibfnamefont {P.}~\bibnamefont {Moras}},
  \bibinfo {author} {\bibfnamefont {S.~K.}\ \bibnamefont {Mahatha}}, \bibinfo
  {author} {\bibfnamefont {C.}~\bibnamefont {Carbone}}, \ and\ \bibinfo
  {author} {\bibfnamefont {E.}~\bibnamefont {Tosatti}},\ }\href@noop {}
  {\bibfield  {journal} {\bibinfo  {journal} {Phys.\ Rev.\ B}\ }\textbf
  {\bibinfo {volume} {91}},\ \bibinfo {pages} {045432} (\bibinfo {year}
  {2015})}\BibitemShut {NoStop}%
\bibitem [{Note3()}]{Note3}%
  \BibitemOpen
  \bibinfo {note} {The linear dispersion in W(110) is explained by charge
  transfer \cite {Braun14}.}\BibitemShut {Stop}%
\bibitem [{\citenamefont {Wang}\ and\ \citenamefont {Johnson}(2011)}]{Wang11}%
  \BibitemOpen
  \bibfield  {author} {\bibinfo {author} {\bibfnamefont {L.-L.}\ \bibnamefont
  {Wang}}\ and\ \bibinfo {author} {\bibfnamefont {D.~D.}\ \bibnamefont
  {Johnson}},\ }\href@noop {} {\bibfield  {journal} {\bibinfo  {journal}
  {Phys.\ Rev.\ B}\ }\textbf {\bibinfo {volume} {83}},\ \bibinfo {pages}
  {241309} (\bibinfo {year} {2011})}\BibitemShut {NoStop}%
\bibitem [{\citenamefont {Franz}\ and\ \citenamefont
  {Molenkamp}(2013)}]{Franz13}%
  \BibitemOpen
  \bibfield  {author} {\bibinfo {author} {\bibfnamefont {M.}~\bibnamefont
  {Franz}}\ and\ \bibinfo {author} {\bibfnamefont {L.~W.}\ \bibnamefont
  {Molenkamp}},\ }\href@noop {} {\emph {\bibinfo {title} {Topological
  Insulators}}},\ edited by\ \bibinfo {editor} {\bibfnamefont {E.}~\bibnamefont
  {Burstein}}, \bibinfo {editor} {\bibfnamefont {A.~H.}\ \bibnamefont
  {MacDonald}}, \ and\ \bibinfo {editor} {\bibfnamefont {P.~J.}\ \bibnamefont
  {Stiles}},\ Vol.~\bibinfo {volume} {6}\ (\bibinfo  {publisher} {Elsevier},\
  \bibinfo {address} {Oxford},\ \bibinfo {year} {2013})\BibitemShut {NoStop}%
\bibitem [{\citenamefont {Rotenberg}\ \emph {et~al.}(1999)\citenamefont
  {Rotenberg}, \citenamefont {Chung},\ and\ \citenamefont
  {Kevan}}]{Rotenberg99}%
  \BibitemOpen
  \bibfield  {author} {\bibinfo {author} {\bibfnamefont {E.}~\bibnamefont
  {Rotenberg}}, \bibinfo {author} {\bibfnamefont {J.~W.}\ \bibnamefont
  {Chung}}, \ and\ \bibinfo {author} {\bibfnamefont {S.~D.}\ \bibnamefont
  {Kevan}},\ }\href@noop {} {\bibfield  {journal} {\bibinfo  {journal} {Phys.\
  Rev.\ Lett.}\ }\textbf {\bibinfo {volume} {82}},\ \bibinfo {pages} {4066}
  (\bibinfo {year} {1999})}\BibitemShut {NoStop}%
\bibitem [{\citenamefont {Shikin}\ \emph {et~al.}(2008)\citenamefont {Shikin},
  \citenamefont {Varykhalov}, \citenamefont {Prudnikova}, \citenamefont
  {Usachov}, \citenamefont {Adamchuk}, \citenamefont {Yamada}, \citenamefont
  {Riley},\ and\ \citenamefont {Rader}}]{Shikin08}%
  \BibitemOpen
  \bibfield  {author} {\bibinfo {author} {\bibfnamefont {A.~M.}\ \bibnamefont
  {Shikin}}, \bibinfo {author} {\bibfnamefont {A.}~\bibnamefont {Varykhalov}},
  \bibinfo {author} {\bibfnamefont {G.~V.}\ \bibnamefont {Prudnikova}},
  \bibinfo {author} {\bibfnamefont {D.}~\bibnamefont {Usachov}}, \bibinfo
  {author} {\bibfnamefont {V.~K.}\ \bibnamefont {Adamchuk}}, \bibinfo {author}
  {\bibfnamefont {Y.}~\bibnamefont {Yamada}}, \bibinfo {author} {\bibfnamefont
  {J.~D.}\ \bibnamefont {Riley}}, \ and\ \bibinfo {author} {\bibfnamefont
  {O.}~\bibnamefont {Rader}},\ }\href@noop {} {\bibfield  {journal} {\bibinfo
  {journal} {Phys.\ Rev.\ Lett.}\ }\textbf {\bibinfo {volume} {100}},\ \bibinfo
  {pages} {057601} (\bibinfo {year} {2008})}\BibitemShut {NoStop}%
\bibitem [{\citenamefont {Chernov}(2015)}]{Chernov15}%
  \BibitemOpen
  \bibfield  {author} {\bibinfo {author} {\bibfnamefont {S.~V.}\ \bibnamefont
  {Chernov}},\ }\href@noop {} {\bibfield  {journal} {\bibinfo  {journal}
  {Ultramicroscopy}\ }\textbf {\bibinfo {volume} {159}},\ \bibinfo {pages}
  {453} (\bibinfo {year} {2015})}\BibitemShut {NoStop}%
\bibitem [{\citenamefont {Thiele}\ \emph {et~al.}(2007)\citenamefont {Thiele},
  \citenamefont {D\"orr}, \citenamefont {Bilani}, \citenamefont {R\"odel},\
  and\ \citenamefont {Schultz}}]{Thiele07}%
  \BibitemOpen
  \bibfield  {author} {\bibinfo {author} {\bibfnamefont {C.}~\bibnamefont
  {Thiele}}, \bibinfo {author} {\bibfnamefont {K.}~\bibnamefont {D\"orr}},
  \bibinfo {author} {\bibfnamefont {O.}~\bibnamefont {Bilani}}, \bibinfo
  {author} {\bibfnamefont {J.}~\bibnamefont {R\"odel}}, \ and\ \bibinfo
  {author} {\bibfnamefont {L.}~\bibnamefont {Schultz}},\ }\href@noop {}
  {\bibfield  {journal} {\bibinfo  {journal} {Phys.\ Rev.\ B}\ }\textbf
  {\bibinfo {volume} {75}},\ \bibinfo {pages} {054408} (\bibinfo {year}
  {2007})}\BibitemShut {NoStop}%
\bibitem [{\citenamefont {B\"ottcher}\ and\ \citenamefont
  {Henk}(2013)}]{Boettcher13}%
  \BibitemOpen
  \bibfield  {author} {\bibinfo {author} {\bibfnamefont {D.}~\bibnamefont
  {B\"ottcher}}\ and\ \bibinfo {author} {\bibfnamefont {J.}~\bibnamefont
  {Henk}},\ }\href@noop {} {\bibfield  {journal} {\bibinfo  {journal} {J.
  Phys.: Condens.\ Matt.}\ }\textbf {\bibinfo {volume} {25}},\ \bibinfo {pages}
  {1360051} (\bibinfo {year} {2013})}\BibitemShut {NoStop}%
\bibitem [{\citenamefont {Bychkov}\ and\ \citenamefont
  {Rashba}(1984{\natexlab{c}})}]{Bychkov84c}%
  \BibitemOpen
  \bibfield  {author} {\bibinfo {author} {\bibfnamefont {Y.~A.}\ \bibnamefont
  {Bychkov}}\ and\ \bibinfo {author} {\bibfnamefont {E.~I.}\ \bibnamefont
  {Rashba}},\ }\href@noop {} {\bibfield  {journal} {\bibinfo  {journal} {Pis'ma
  Zh.\ Eksp.\ Teor.\ Fiz.}\ }\textbf {\bibinfo {volume} {39}},\ \bibinfo
  {pages} {66} (\bibinfo {year} {1984}{\natexlab{c}})},\ \bibinfo {note}
  {translation in Ref.~\cite{Bychkov84a}}\BibitemShut {NoStop}%
\end{thebibliography}%
\end{document}